\documentclass[%
 reprint,
 amsmath,amssymb,
 aps,
]{revtex4-2}

\usepackage{graphicx}%
\usepackage{dcolumn}%
\usepackage{bm}%

\PassOptionsToPackage{table}{xcolor}
\usepackage{orcidlink}
\usepackage[capitalise]{cleveref}
\usepackage{aasmacros}
\usepackage{graphicx} 
\usepackage[utf8]{inputenc} 
\usepackage{verbatim}
\usepackage[normalem]{ulem}
\usepackage{mathtools}

\usepackage[commandnameprefix=ifneeded, authormarkup=none,draft]{changes}

\newcommand{\Ins}{\affiliation{Dipartimento di Scienza e Alta Tecnologia, Università dell’Insubria, via Valleggio 11, I-22100 Como, Italy}}
\newcommand{\Clap}{\affiliation{Como Lake centre for AstroPhysics (CLAP), DiSAT, Università dell’Insubria, via Valleggio 11, 22100 Como, Italy}}
\newcommand{\Bic}{\affiliation{Dipartimento di Fisica “G. Occhialini”, Università degli Studi di Milano-Bicocca, Piazza della Scienza 3, 20126 Milano, Italy}}
\newcommand{\Infn}{\affiliation{INFN, Sezione di Milano-Bicocca, Piazza della Scienza 3, 20126 Milano, Italy}}
\newcommand{\bham}{\affiliation{Institute for Gravitational Wave Astronomy \& School of Physics and
Astronomy, University of Birmingham, Birmingham, B15 2TT, UK}}
\newcommand{\aei}{\affiliation{Max Planck Institute for Gravitational Physics (Albert Einstein Institute), Am M\"uhlenberg 1, Potsdam 14476, Germany}}
\newcommand{\esa}{\affiliation{European Space Agency (ESA), European Space Research and Technology Centre (ESTEC), Keplerlaan 1, 2201 AZ Noordwijk, the Netherlands}}

\definechangesauthor[name={RB}, color=teal]{rb} %

\newcommand{\dd}{{\mathrm d}}

\begin{document}
\title{Is your stochastic signal really detectable?}

\author{Federico~Pozzoli\orcidlink{0009-0009-6265-584X}} 
\Ins
\Infn
\Clap
\email{fpozzoli@uninsubria.it}
\author{Jonathan~Gair\orcidlink{0000-0002-1671-3668}}\aei
\author{Riccardo~Buscicchio\orcidlink{0000-0002-7387-6754}}
\Bic 
\Infn
\bham 
\author{Lorenzo Speri\orcidlink{0000-0002-5442-7267}}\esa

\begin{abstract}
Separating a stochastic gravitational wave background (SGWB) from noise is a challenging statistical task. 
One approach to establishing a detection criterion for the SGWB is using Bayesian evidence. If the evidence ratio (Bayes factor) between models with and without the signal exceeds a certain threshold, the signal is considered detected. We present a formalism to compute the averaged Bayes factor, incorporating instrumental-noise and SGWB uncertainties. As an example, we consider the case of power-law-shaped SGWB in LISA and generate the corresponding \textit{bayesian sensitivity curve}. Unlike existing methods in the literature, which typically neglect uncertainties in both the signal and noise, our approach provides a reliable and realistic alternative. This flexible framework opens avenues for more robust stochastic gravitational wave background detection across gravitational-wave experiments.
\end{abstract}

\maketitle

\section{Introduction}
Stochastic signals arising from the incoherent superposition of numerous faint sources generated by some physical process are the target for many different types of scientific experiment. One example 
is the stochastic gravitational wave background (SGWB) sourced by either astrophysical systems or cosmological processes. 
At nHz frequencies, the first evidence for a SGWB has recently been reported by multiple collaborations~\cite{2022MNRAS.510.4873A,2023A&A...678A..50E,2023ApJ...951L...6R,2023ApJ...951L...8A}.
SGWB inference presents a number of challenges, ubiquitous across the spectrum. In particular, it entails disentangling the contribution from the target physical process from that of the instrumental noise, about which knowledge is usually limited, and often requires distinguish the contributions from multiple overlapping components.

Several features can be leveraged to improve the ability of a detector to resolve a SGWB. Spectral signatures, as well as non-stationarities and source anisotropies are among the most explored.
In order to assess whether a certain stochastic signal is detectable a notion of statistical significance has to be identified. Ref.~\cite{2013PhRvD..88l4032T} introduced the concept of \textit{power-law sensitivity} (PLS), a convenient graphical method to construct a sensitivity curve for power-law-shaped stochastic spectra, specifically SGWB observed by current or future detectors.
The signal-to-noise ratio ($\mathrm{SNR}$) is at the very core of the PLS-based detectability:  any power-law crossing the PLS corresponds to a SGWB observable with an expected SNR above a given threshold. 
Typically, the SNR is defined as the ratio between the expected value and the root-mean-square of a statistic (e.g., the zero-lag cross-correlation of two detectors) in the presence and  absence of signal, respectively \cite{1999PhRvD..59j2001A}. 
However, this definition corresponds to a statistic assuming perfect knowledge on both the signal and noise spectral shapes and noise level, neglecting their respective prior uncertainties.
Several studies in the literature have focused on relaxing such assumptions~\cite{2020CQGra..37u5017K, 2023JCAP...04..066B,2024PhRvD.109d2001M}.

In this manuscript, we propose a novel formalism that systematically addresses these limitations. 
We use the Bayes factor as our fundamental metric to compare the evidences of two competing models and suggest an approach to rapidly evaluate model evidences with and without a SGWB component in the presence of noise, while marginalising over uncertainties in both. 
Then, following the standard PLS methodology, we identify numerically the envelope for the family of signals yielding a desired Bayes factor, hence defining the Bayesian equivalent of the PLS. Henceforth, we will refer to it as \emph{Bayesian power-law sensitivity} (BPLS). The BPLS has the same applicability of the PLS, though it fully accounts for marginalisation over model uncertainties.
As a practical example, we apply our method to the detection of SGWBs with the Laser Interferometer Space Antenna (LISA) \cite{2024arXiv240207571C}: 
the dubious availability ---in realistic scenarios--- of noise-informative but GW-insensitive data combinations, often referred to as null-channels~\cite{2022PhRvD.105b3009M}, makes it a compelling approach because of the crucial role played by model uncertainties.
As expected, larger uncertainties yield smaller significances at fixed SGWB amplitudes. 
This is a central result, as it informs more realistically figures of merit on LISA SGWB reconstructions.

\section{Modeling}
As previously mentioned, there are features that can be incorporated into the analysis to better characterize a stochastic signal and enhance detection. Throughout this paper, we assume that both signal and noise are realizations of independent, Gaussian, and stationary processes, representing the easiest yet 'worst-case' scenario for studying the detectability of a stochastic signal.
We denote with $\mathbf{d} = \{\mathbf{d}_1, \dots, \mathbf{d}_p\}$ the datastreams of $p$ detectors in the time domain. These datastreams can represent outputs from a variety of detectors, which for gravitational wave detectors could be Time-Delay Interferometry (TDI) channels from space-based interferometers \cite{2021LRR....24....1T}, pulsar timing residuals \cite{2018CQGra..35m3001V}, or individual measurements from ground-based detectors \cite{2021PhRvD.104b2004A}.
We divide the total observation time into $n_c$ adjacent segments, each of duration $T = T_{\mathrm{obs}} / n_c$. 
We assume the datastreams to be uniformly sampled, with a cadence $\Delta t$, and denote each segment by ${\mathbf{d}}_{(c)}=(\mathbf{d}_{(c){{1}}},\dots,\mathbf{d}_{(c){{p}}})$.
We use the segments to construct coarse-grained data, and define the averaged periodograms in the frequency domain as
\begin{align}
    \overline{P}(f) = \frac{1}{n_c}\sum_{c=1}^{n_c}\tilde{\mathbf{d}}_{(c)}(f)\tilde{\mathbf{d}}_{(c)}^\dagger(f),
    \label{eq:periodogram}
\end{align}
where $\tilde{\mathbf{d}}_{(c)}$ denotes the Fourier transform of the $c-$th segment, and $\tilde{\mathbf{d}}^\dagger_{(c)}$ its transpose conjugate.
Following this approach, we reduce the computational cost,  but also limit the low-frequency resolution. 
An upper-limit on $n_c$ is imposed if we want to preserve information at the lowest frequencies, $n_c < f_{\rm min} T_{\rm obs}$, where $T_{\rm obs}$ denotes the original duration of the datastream and $f_{\rm min}$ is the lowest frequency we want to resolve.

Under Gaussianity assumption, the matrix $Y(f) = \overline{P}(f)$ follows a complex Wishart distribution with $n_c$  degrees of freedom and scale matrix 
\begin{align}
 \Gamma(f) = \frac{1}{n_c}\left(\Sigma_n(f) + \Sigma_{\mathrm{GW}}(f)\right), 
 \label{eq:totalcovariance}
\end{align}
where $\Sigma_n$ and $\Sigma_{\mathrm{GW}}$ denote the noise and GW signal covariance matrices, respectively.
A lower-limit on $n_c$ must also be enforced: $n_c > p - 1$, a defining property of the complex Wishart distribution. In the opposite regime, data in full frequency resolution can be employed, accounting for the multivariate normal distribution describing their likelihood.  
The complex Wishart likelihood at a given frequency reads
\begin{equation}
    \!\mathcal{L}(Y(f) | \Gamma(f)) \!\propto \!\!\frac{|Y(f)|^{n_c - p }}{|\Gamma(f)|^{n_c}}\exp\left[-\mathrm{tr}(\Gamma^{-1}(f)Y(f))\right] 
    \label{eq:likelihood_bin}
\end{equation}
where $\text{tr}$ and $|\cdot|$ denote respectively the trace and the determinant operator.
To evaluate model evidences, priors must be specified and marginalized over as follows
\begin{align}
      {\cal Z}(Y) =  \int \dd \Gamma {\cal L}(Y \mid \Gamma) \pi(\Gamma),
    \label{eq:evidence}  
\end{align}
where we drop the frequency dependence of the matrices for brevity.
We choose priors conjugate to the likelihood in~\cref{eq:likelihood_bin} for two main reasons: first, upon inference the posterior preserves the same functional form of the prior; second, as we shall see below, the free parameters available yield enough flexibility to choose a reference expectation value and variance for $\Gamma$.
We assume a complex inverse-Wishart prior $\mathcal{CW}^{-1}$ on \( \Gamma \), with \( \nu \) degrees of freedom and scale matrix \( \Psi \). 
\begin{align}
    \Gamma \mid \Psi, \nu &\sim {\cal CW}^{-1}(\Psi, \nu)\\
    \pi (\Gamma| \Psi, \nu) &\propto |\Psi|^{\nu}|\Gamma|^{-(\nu + p)} \exp{\left[-\mathrm{tr}(\Psi\Gamma^{-1})\right]}.
\end{align}
The chosen prior yields an expectation value for $\Gamma$
\begin{equation}
    E[\Gamma] = \frac{\Psi}{(\nu -p)}  \label{eq:expectation}
\end{equation} 
for $\nu > p+1$. 
Therefore, we set the scale-matrix parameter of the prior to 
\begin{align}
    \Psi = \frac{\nu - p}{n_c}\left(\Sigma_n^0 + \Sigma_{\rm GW}(\vec{\lambda})\right),
    \label{eq:scale_prior}
\end{align}
where $\Sigma_n^0$ denotes a fiducial estimate of the noise covariance matrix, and $\vec{\lambda}$ are the parameters describing the SGWB power spectral density (e.g., the amplitude and slope for a power-law signal). As a consequence, the mean of the prior distribution is the expected $\Gamma$ in~\cref{eq:expectation}.
The value of $\nu$ can be adjusted to represent the scale of our prior uncertainty on $\Gamma$, since the variances $\mathrm{Var(}\Gamma_{ii})$ are proportional to $\Psi^2_{ii} / \left[(\nu -p)^2(\nu -p -1)\right]$. For instance, $\nu\sim10$ represents almost $10\%$ uncertainty for $p=3$. 

In realistic data-analysis scenarios, the parameter $\vec{\lambda}$ must be inferred, hence a prior on it must be placed. On the contrary following the PLS construction, here we will explore it systematically.  
The posterior distribution for $\Gamma$ follows a complex inverse-Wishart distribution with parameters updated according to the observations, which are described by the amount of data observed, $n_c$, and the average periodogram, $Y$, from~\cref{eq:periodogram}:
\begin{align}
    \Gamma \mid Y, \Psi, \nu &\sim \mathcal{CW}^{-1}(\Psi + Y, \nu + n_c),\\
    p(\Gamma|Y, \Psi, \nu) &\!\propto\! \frac{|\Psi|^{\nu}|Y|^{n_c - p}}{|\Gamma|^{n_c + \nu + p}}\exp\!{\left[-\mathrm{tr}((Y + \Psi)\Gamma^{-1})\right]}
    \label{eq:posterior}
\end{align}
Upon marginalization, the evidence in Eq.~\ref{eq:evidence} reads
\begin{align}
    \mathcal{Z}(Y|\Psi,\nu) = C\frac{|\Psi|^\nu|Y|^{n_c - p}}{|\Psi + Y|^{(\nu +n_c)}},
\end{align}
where $C$ is a suitable normalization constant, independent on $\Psi,\nu$.
Notably, the evidence retains a dependence on the prior hyperparameters $\Psi$ and $\nu$.
Reinstating explicitly the dependence on the frequency, the Bayes factor for a given bin---between the presence ($H_1$) and absence ($H_0$) of GW signal hypotheses--- can be written as:

\begin{align}
    \log\mathcal{B}(f|\nu, n_c, \vec\lambda, Y) 
    & ={(n_c + \nu)}\log\left|\frac{\Psi(f|H_0) + Y(f)}{\Psi(f|H_1) + Y(f)}\right| +\nonumber \\ 
    & + \nu \log\left|\frac{\Psi(f|H_1)}{\Psi(f|H_0)}\right| ,\label{eq:bayes_factor}
\end{align}
where the scale matrices under the two hypotheses read:
\begin{align}
    &\Psi(f|H_1) = \frac{\Sigma_n^0(f) + \Sigma_{\mathrm{GW}}(f, \vec{\lambda})}{n_c}(\nu - p ), \\
    &\Psi(f|H_0) = \frac{\Sigma_n^0(f)}{n_c}(\nu - p ).
\end{align}
Under the assumption that the signal and noise are stationary, the overall Bayes factor is obtained by taking the product over all frequencies.
\cref{eq:bayes_factor} is the central result of this work, as it allows for quick evaluation of Bayes factors assuming \emph{a priori} uncertain knowledge on the noise and signal. 
From~\cref{eq:bayes_factor}, we now want to identify the values of $\vec{\lambda}$ yielding a specified Bayes factor. Typically, we are interested in the contour corresponding to $\log(\mathcal{B}) \geq 1$ as fiducial threshold of strong evidence in favor of $H_1$.

Since \cref{eq:bayes_factor} depends on the specific observed realization of $Y$, we compute the expectation value of $\log(\mathcal{B})$ over multiple realisations. Two viable approaches are possible here: numerical and analytical approximation. 
For the former, given each $\vec{\lambda}$, we draw realizations of $Y$ from the Wishart distribution and compute the average of the logarithm in~\cref{eq:bayes_factor}.
For the latter, we first rewrite Eq.~\ref{eq:bayes_factor} as
\begin{align}
    \ln\mathcal{B} &= \nu \mathrm{tr}(\ln \Psi_1 - \ln\Psi_0) +  \nonumber\\
    & +(n_c + \nu)\mathrm{tr}(\ln(\mathbb{I} + Y^{-1}\Psi_0) + \nonumber \\
    & - \ln(\mathbb{I} + Y^{-1}\Psi_1)).
\end{align}
using the identity $\ln |M| = \mathrm{tr}(\ln M)$. We now use the expansion $\ln \left(\mathbb{I} + Y^{-1}\Psi) = (\sum_{k = 1}^\infty(-1)^{k+1} (Y^{-1}\Psi)^k/k\right)$, which holds under the assumption that  $ Y^{-1}\Psi \ll 1 $. Since $\Psi \propto (\nu - p)/ n_c$, this condition corresponds to the limit where $ \nu - p $ is small and $n_c$ is large. Practically, this represents a scenario characterized by significant uncertainty in the process and a long observation period. At leading order, the log Bayes factor reads 
\begin{align}
    \label{eq:first_order}
    \ln \mathcal{B} & = \nu \mathrm{tr}(\ln \Psi_1 - \log \Psi_0) +  \nonumber\\
    &+\!(\nu \!+\! n_c)\mathrm{tr}(Y^{-1}\Psi_1\! -\! Y^{-1}\Psi_0)\!+\!\mathcal{O}(Y^{-1}\Psi),
\end{align}
and taking its expectation value
\begin{align}
    \label{eq:first_order2}
    \!\!\!\!\!E[\ln \mathcal{B}] & = \nu \mathrm{tr}(\ln \Psi_1 - \ln \Psi_0) +  \nonumber\\
    &+\!\frac{(\nu \!+\! n_c)}{n_c -p}\mathrm{tr}(\Gamma^{-1}\Psi_1\! -\! \Gamma^{-1}\Psi_0)\!+\!\mathcal{O}(Y^{-1}\Psi)\\
    &\approx \nu \mathrm{tr}\left[\ln(\Sigma_n^0 + \Sigma_{\mathrm{GW}}) - \ln\Sigma_n^0\right]+\nonumber\\
    &+\frac{(\nu + n_c)(\nu-p)}{n_c - p}\mathrm{tr}\!\left[(\Sigma_n^0 + \Sigma_{\mathrm{GW}})^{-1}\Sigma_n^0 - \mathbb{I}\right] \label{eq:expanded_expectation2}
\end{align}
where we used the relation $E\mathrm{[Y^{-1}]} = \Gamma^{-1} / (n_c- p)$ \cite{Vonrosen1988}. 
Even though~\cref{eq:first_order} can be expanded by including higher order moments from the complex inverse-Wishart distribution~\cite{Maiwald1997}, we provide here results at leading-order only. As expected in the case of $\Sigma^0 _n \gg \Sigma_{\rm GW}$, the expectation value of the log-Bayes factor goes to 0. Instead for $\Sigma^0 _n \ll \Sigma_{\rm GW}$, the expectation value is
\begin{align}
    \nu \mathrm{tr}\left[\ln( \Sigma_{\mathrm{GW}} (\Sigma_n^0)^{-1}) \right]
 - \frac{(\nu + n_c)(\nu-p)}{n_c - p}\mathrm{tr}\!\left[ \mathbb{I}\right] \, .
\end{align}
\section{Results}
We show here a concrete example by considering the SGWB detection problem for LISA. Specifically, we consider a four-year mission duration. To reduce the computational cost we restrict our analysis to the frequency range $\left[10^{-4}, 10^{-2} \right] \rm Hz$. 
We assume three datastreams, i.e., $p=3$, each corresponding to a TDI variable. TDI is a post-processing technique that reduces the laser frequency noise in the LISA measurements.
Various combinations of TDI variables can be constructed under different approximations to the orbital motion of the LISA satellites \cite{2021LRR....24....1T}.
\begin{figure*}[t!]
    \centering    \includegraphics[width=2\columnwidth]{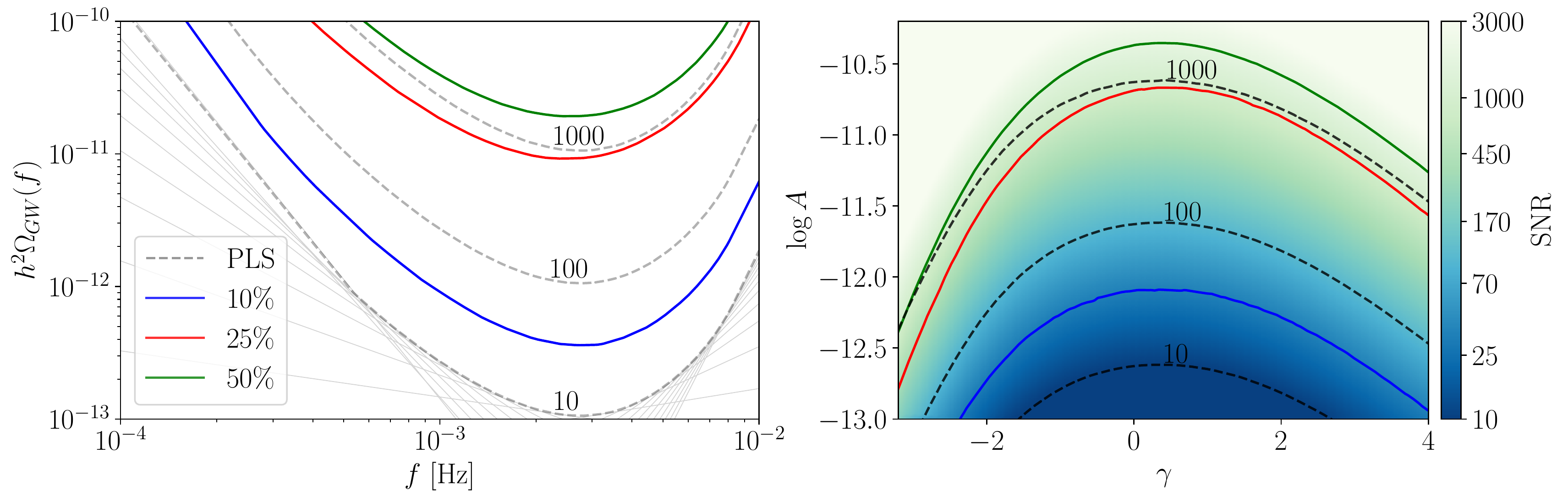}
    \caption{SGWB power-law sensitivities with Bayes factor threshold of 10, for three reference noise uncertainties $(10\%,25\%,50\%)$. (\textit{Left panel}) Bayesian power-law sensitivity corresponding to a log-Bayes factor threshold of one. The colors represent different numbers of prior degrees of freedom, $\nu$, reflecting varying levels of uncertainty. The dashed curves represent the power-law sensitivity for different SNR thresholds ($10$, $100$, $1000$). For reference, we show as solid light grey lines the powerlaw spectra used to construct the PLS at SNR 10.(\textit{Right panel}) The solid curves denote contours corresponding to $E\left[\log(\mathcal{B})\right] = 1$ in the parameter space of the amplitude and slope of a power-law spectrum. The dashed lines represent contours of fixed SNR within the same parameter space.
    }
    \label{fig:bayesian_pls}
\end{figure*}
\begin{figure}[t!]
    \centering
    \includegraphics[width=\columnwidth]{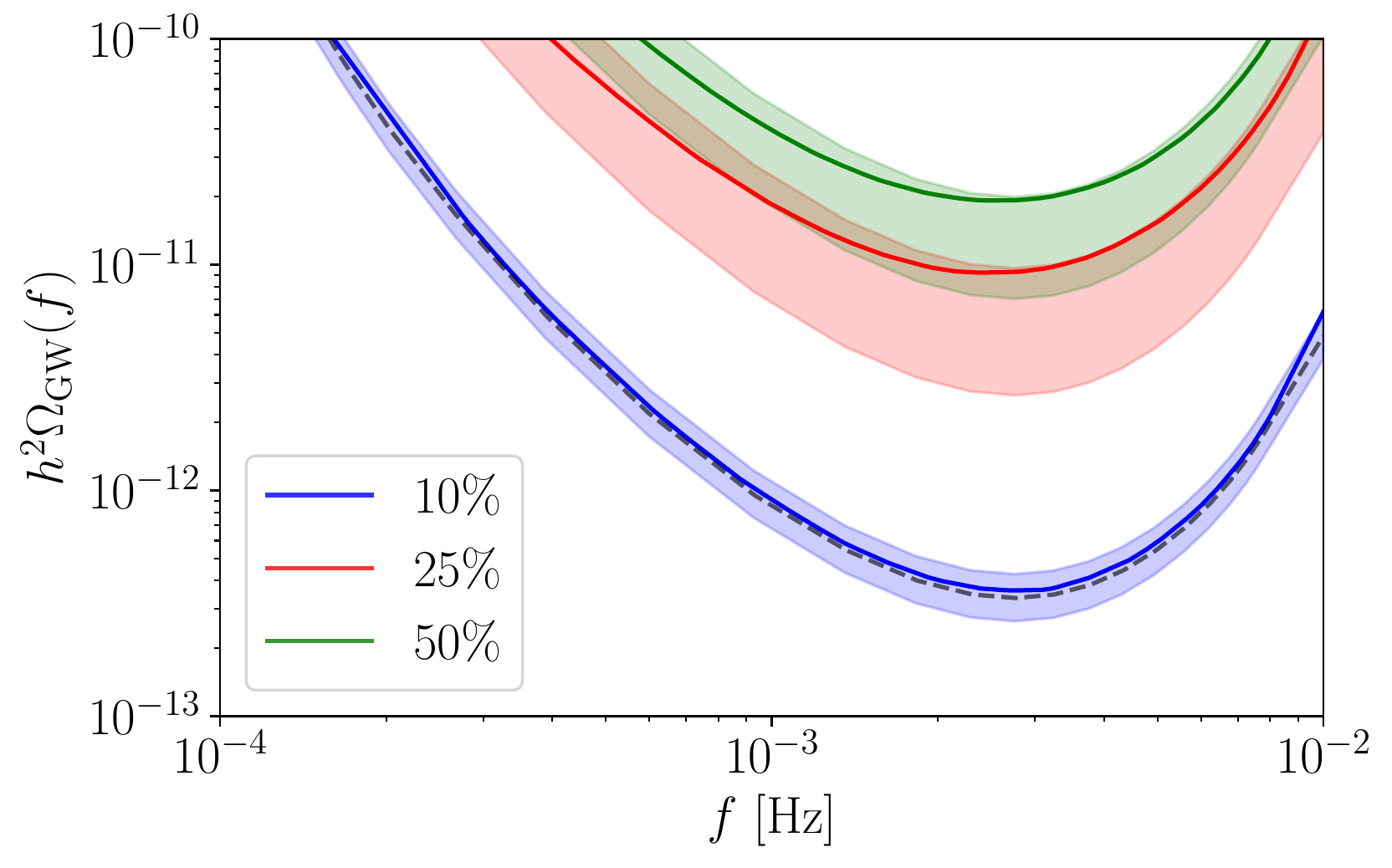} 
    \caption{Mapping between BPLS and standard PLS.
    The solid curves represent the same BPLS curves shown in the left panel of ~\cref{fig:bayesian_pls}.
    The shaded region encompasses the PLS curves intersecting the BPLS at least once. 
    Therefore, SNRs associated with the PLS curves in each band correspond to SGWB signals detectable at $\log \mathcal{B} = 1$ or higher. 
    For reference, the dashed curve denotes a PLS closely matching the BPLS associated with $\nu_{10}$, and corresponds to an ${\rm SNR}=35$ over $4 {\rm yr}$ of observation. Each color corresponds to a different level of a priori uncertainties, as described by the parameter $\nu$.}
    \label{fig:bayesian_pls1_1}
\end{figure}
In this work, we assume equal and constant armlengths, hence use the $A,E,T$ variables as the datastreams~\cite{2021LRR....24....1T}. In doing so, the model for $\Gamma$ simplifies to a diagonal matrix. 
Under these assumptions, the SNR of a signal with parameters $\vec\lambda$ reads
\begin{align}
\label{eq:snr}
    {\rm SNR}^2\left(\vec\lambda\right)\!\!=T_{\rm obs}\!\!\!\!\sum_{i = \left\{A, E, T\right\}}\int\displaylimits_{0}^{\infty}\!\!\dd f \left(\frac{R_i(f)S_h(f, \vec\lambda)}{P_{n, i}(f)}\right)^2,
\end{align}
where $R_i$ is the sky-averaged response function of LISA to SGWB ---summed over the two independent polarizations--- and $P_{n,i}$ is the power spectral density (PSD) for the noise in the $i$-th TDI channel.  
Analytical expressions for $R_i$ and $P_{n,i}$ are available in the literature~\cite{2023PhRvD.107l3531H, 2019CQGra..36j5011R, 2021arXiv210801167B}. 
\begin{figure*}
    \centering    
    \includegraphics[width=2\columnwidth]{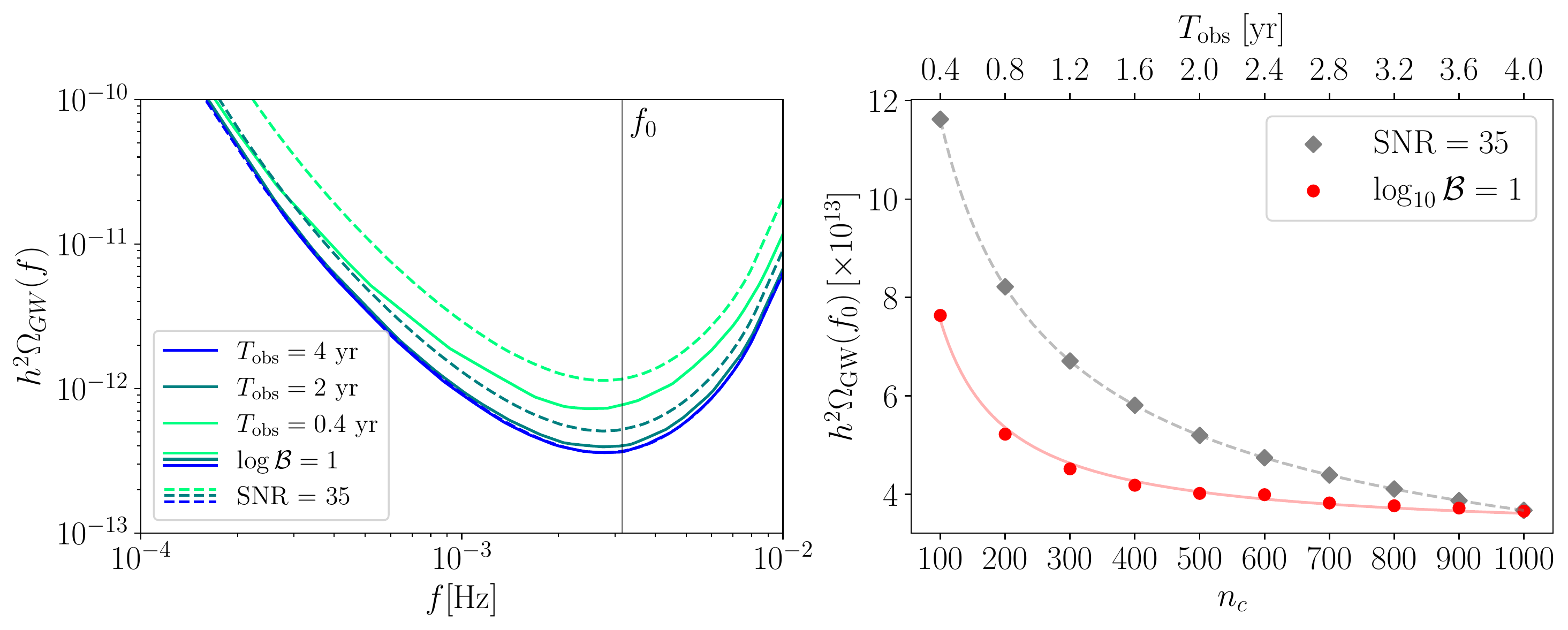}
    \caption{Dependency of BPLS on the observing time. (\textit{Left panel}) The dashed lines refer to PLSs constructed for a target $\mathrm{SNR} = 35$, while solid ones denote the BPLSs at $\log\mathcal{B} = 1$ considering $10\%$ prior uncertainties in the covariance matrix. Colors denote three fiducial observation times, $T_{\rm obs} = 0.4{\rm yr}, 2{\rm yr}, 4 {\rm yr}$, as indicated in the  legend.
    (\textit{Right panel}) Dependency of the PLS (grey) and BPLS (red) value at a reference frequency $f_0 = 10^{-2.5}\mathrm{Hz}$ as a function of the number of chunks, i.e., the observation time. Dashed (solid) curve represent the best fit of 10 PLS (BPLS) values with as expected from Eq.~\eqref{eq:snr} and Eq.~\eqref{eq:expanded_expectation2}, respectively.}
    \label{fig:bayesian_pls2}
\end{figure*}
It is important to note that Eq.\ref{eq:snr} is valid in the regime where the noise dominates over the signal. Otherwise, more robust statistics need to be considered as in \cite{2015MNRAS.451.2417R}.

SGWBs are often characterized by $\Omega_{\mathrm{GW}}$, the GW energy density per logarithmic frequency. This is related to the PSD via:
\begin{align}
    \label{eq:spectral_density}
    S_{h}(f, \vec{\lambda}) = \frac{3H_0^2}{4\pi^2f^3}\Omega_{\mathrm{GW}}(f,\vec{\lambda}),
\end{align}
where $H_0$ is the Hubble parameter at the present day.
We consider power-law shaped spectra, characterized by their amplitude $A=\Omega_{\rm GW}(f = 10^{-2.5} \mathrm{Hz})$ and slope $\gamma$:
\begin{align}
\Omega_{\mathrm{GW}}(f, \vec{\lambda} = \left(A, \gamma\right)) = A\left(\frac{f}{10^{-2.5}\mathrm{Hz}}\right)^\gamma.
\end{align}

We evaluate the expectation value of the Bayes factor as a function of $(A,\gamma)$, for three reference values of $\nu$: $\nu_{10}=10$, $\nu_{25}=8$, and $\nu_{50}=7$, corresponding to $10\%$, $25\%$ and $50\%$ prior uncertainties on the covariance in~\cref{eq:totalcovariance}.

For each $\nu$, we construct the contour for $\vec{\lambda}$ corresponding to the level $E\left[\log(\mathcal{B})\right] = 1$. This is shown in the right panel of~\cref{fig:bayesian_pls}.
The contour $(A(s), \gamma(s))$ describes a collection of power-law signals detectable with a fixed, Bayesian significance. 
By identifying at each frequency the one yielding the highest $\Omega_{\rm GW}(f)$, we define the envelope $(A_{\nu}(f), \gamma_{\nu}(f))$ 
\begin{align}
    \Omega_{\rm GW}^{\mathrm{BPLS}}(f) & = A_\nu(f)\left(\frac{f}{10^{-2.5}\mathrm{Hz}}\right)^{\gamma_{\nu}(f)}\\
    &\coloneqq \max_s\left[A(s)\left(\frac{f}{10^{-2.5}\mathrm{Hz}}\right)^{\gamma(s)}\right]\label{eq:omega_BPLS}.
\end{align}
This is the \textit{Bayesian power-law Sensitivity} (BPLS). We show the resulting curve in the left panel of~\cref{fig:bayesian_pls}.
For a direct comparison, we over-plot the standard SNR-based PLS, following Ref.~\cite{2013PhRvD..88l4032T}. 

An unambiguous mapping between PLS and BPLS is possible only in the regime where the posterior distribution is dominated by the likelihood, such that $\log(\mathcal{B}) \propto \mathrm{SNR}^2$\cite{2011PhRvD..84f2003C}. In general, such a comparison is non-trivial. Notably, we find that the relationship between the two quantities becomes increasingly degenerate as the prior uncertainty in the covariance matrix grows. 
We illustrate this effect in~\cref{fig:bayesian_pls1_1}. 
Therein, we identify the set of PLSs (shown as shaded areas) intersecting a given BPLS at least once.
Each intersection corresponds to an SGWB signal yielding a certain SNR and $\log\mathcal{B} = 1$. 
As shown in the figure, to higher prior uncertainties correspond larger PLS bands, hence a larger set of SNRs.
SNRs yielding a detection significance $\log{\cal B} = (0.5, 1, 2)$ ---substantial, strong, and decisive, respectively--- may range in $([20,30], [25,40], [37,60])$ for $\nu_{10}$, in $([248,917], [253,920], [263,924])$ for $\nu_{25}$, and in $([664,1897], [673,1900], [690,1905])$ for $\nu_{50}$.

In the analysis presented so far, we assumed a nominal mission duration of four years \cite{2024arXiv240207571C}. To explore the dependence of the BPLS on the observation time, we vary $T_{\mathrm{obs}}$ by adjusting the number of chunks $n_c$ at fixed chunk duration, while keeping the prior uncertainty level fixed at $\nu_{10}$.
We focus on the dependency on $T_{\rm obs}$ of the BPLS at a single significance $\log \mathcal{B} = 1$, which closely matches the PLS with $\mathrm{SNR} = 35$ over the nominal mission duration $T_{\mathrm{obs}} = 4 \mathrm{yr}$. 
We illustrate our findings in the left panel of~\cref{fig:bayesian_pls2}. 
Notably, the PLS shows a stronger dependency on time than the BPLS.
At fixed frequency and SNR, the PLS level is inversely proportional to the square root of the observation time, i.e., $\propto 1/\sqrt{T_{\mathrm{obs}}}$.
Instead, the BPLS depends non-trivially on $T_{\rm obs}$: from the expansion in~\cref{eq:expanded_expectation2} and considering  $n_c = T_{\mathrm{obs}} / T$, we observe that at leading order $E\left[\ln \mathcal{B}\right] \propto (\nu + n_c) / (n_c - p) = (\nu T + T_{\mathrm{obs}})/(T_{\mathrm{obs}} - p T)$.
A fit of this form agrees well with the numerical data (see right panel of~\cref{fig:bayesian_pls2}). The BPLS does in fact show the expected $1/\sqrt{T_{\mathrm{obs}}}$ behaviour, but only in the regime $n_c \ll \nu$. This can be derived by considering the $\nu \rightarrow \infty$ limit of Eq.~(\ref{eq:bayes_factor}). The physical interpretation is that we are able to measure the total stochastic component in the data with a certain precision, which increases as the number of segments increases. 
For small observation times, the uncertainty in the measured background is large compared to the prior uncertainty and so the latter is irrelevant: the BPLS behaves like the PLS in this regime. As the number of segments increases the precision on the total stochastic component of the data improves and there is a transition into a regime in which the prior uncertainty dominates. From this point onwards the background constraint can no longer improve: at leading order, the background level observable at a fixed $\cal B$ exhibits a saturation effect.

\section{Conclusion}
Inferring the spectral shape of an SGWB is a non-trivial task.
However, in the idealized case of power-law signals, their detectability can be conveniently cast in a simple form.
In particular, we employed Bayesian evidence as a detectability metric. By leveraging prior conjugacy, we naturally incorporated noise uncertainty and marginalized over them. In analogy with Ref.~\cite{2013PhRvD..88l4032T}, we developed a robust method to efficiently compute the BPLS.
This curve identifies detectable signal up to a given Bayes factor threshold. To facilitate usage of our results, data behind all figures are available at Ref~\cite{datarelease}.

We applied our method to the LISA problem under idealized mission conditions.
Considering more realistic scenarios, such as correlations between datastreams, introduces additional complexity to the analysis.
Nonetheless, such complexity, and possibly parameterized model uncertainties across the spectrum, is expected improve the SGWB detectability.
However, the proposed approach remains robust even when a non-diagonal covariance matrix is considered.
The BPLS can be extended directly to other signal shape. In that case, we have to construct the $n-1$-dimensional surface at a given Bayes factor, where $n$ denotes the number of parameters describing the template. We then define the Bayesian sensitivity curve by taking the maximum of the PSD over this surface at each frequency.
The proposed methodology is versatile and can be extended to other GW observatories and contexts where stochastic signals are targeted, making it a powerful and flexible tool for the SGWB community.
\begin{acknowledgments}
Data supporting this paper are publicly released at Ref.~\cite{datarelease}.
The authors thank C.~J.~Moore, and A.~Sesana for useful comments.
R.B. is supported by MUR Grant ``Progetto Dipartimenti di Eccellenza 2023-2027'' (BiCoQ),
and the ICSC National Research Centre funded by NextGenerationEU.
R.B. is supported by Italian Space Agency Grant ``Phase A activities for the LISA mission''
No.~2017-29-H.0.
Computational work was performed at Bicocca's Akatsuki cluster (B Massive funded).
\end{acknowledgments}
\nocite{*}
\bibliographystyle{apsrev4-2}
\bibliography{apssamp}
\vfill
\end{document}